%% file: main.tex
\documentclass[conference]{IEEEtran}
\input{preamble}

\input{macros}
\ifCLASSINFOpdf
\else
\fi

\begin{document}

%
\title{Joint Channel Selection using FedDRL in V2X}

\author{\IEEEauthorblockN{Lorenzo Mancini\IEEEauthorrefmark{1}, Safwan Labbi\IEEEauthorrefmark{1},  Karim Abed Meraim\IEEEauthorrefmark{2}, Fouzi Boukhalfa\IEEEauthorrefmark{3}, Alain Durmus\IEEEauthorrefmark{1}, \\ Paul Mangold\IEEEauthorrefmark{1} and Eric Moulines\IEEEauthorrefmark{1}}
 \IEEEauthorblockA{\IEEEauthorrefmark{1}CMAP, Ecole Polytechnique, Paris, France}
 \IEEEauthorblockA{\IEEEauthorrefmark{2}PRISME Laboratory, University of Orléans, Orléans, France }
 \IEEEauthorblockA{\IEEEauthorrefmark{3}Technology Innovation Institute, 9639 Masdar City, Abu Dhabi, UAE \\
    Emails: \{lorenzo.mancini, safwan.labbi, alain.durmus, paul.mangold, eric.moulines\}@polytechnique.edu , \\ karim.abed-meraim@univ-orleans.fr, Fouzi.Boukhalfa@tii.ae
 }}


%


\IEEEoverridecommandlockouts

\IEEEpubid{\hspace{-0.5\textwidth}\begin{minipage}{0.5\textwidth}\ \\[12pt]
  \copyright 2024 IEEE. Personal use of this material is permitted. Permission 
    from IEEE must be obtained for all other uses, in any current or future 
    media, including reprinting/republishing this material for advertising or 
    promotional purposes, creating new collective works, for resale or 
    redistribution to servers or lists, or reuse of any copyrighted 
    component of this work in other works.
\end{minipage}}

\maketitle

\IEEEpubidadjcol
\begin{abstract}
Vehicle-to-everything (V2X) communication technology is revolutionizing transportation by enabling interactions between vehicles, devices, and infrastructures. This connectivity enhances road safety, transportation efficiency, and driver assistance systems. V2X benefits from Machine Learning, enabling real-time data analysis, better decision-making, and improved traffic predictions, making transportation safer and more efficient.

In this paper, we study the problem of joint channel selection, where vehicles with different technologies choose one or more Access Points (APs) to transmit messages in a network.
In this problem, vehicles must learn a strategy for channel selection, based on observations that incorporate vehicles' information (position and speed), network and communication data (Signal-to-Interference-plus-Noise Ratio from past communications), and environmental data (road type). 
We propose an approach based on Federated Deep Reinforcement Learning (FedDRL), which enables each vehicle to benefit from other vehicles' experiences.  
Specifically, we apply the federated Proximal Policy Optimization (FedPPO) algorithm to this task. We show that this method improves communication reliability while minimizing transmission costs and channel switches.
The efficiency of the proposed solution is assessed via realistic simulations, highlighting the potential of FedDRL to advance V2X technology.
\end{abstract}


%
\IEEEpeerreviewmaketitle

\input{introduction}

\input{related-work}

\input{background}

\input{channelselection}

\input{conclusion}


\input{acknowledgement}



%
\printbibliography

\end{document}

%% file: preamble.tex
\usepackage{amsthm}
\usepackage{amsmath}

\usepackage{newtxtext}       %
\usepackage[varvw]{newtxmath}       
\usepackage{newtxtext,newtxmath} 

\usepackage[size=tiny]{todonotes}
\usepackage{url}
\usepackage{comment}
\usepackage{xparse} 
\usepackage{booktabs}
\usepackage{pifont}
\usepackage{xargs}
\usepackage{xstring}

\usepackage[T1]{fontenc}

\usepackage{flushend} 
\usepackage[noend]{algpseudocode}
\usepackage{algorithm}
\usepackage{algorithmicx}
\usepackage[inline]{enumitem}

\usepackage[
backend=biber,
style=numeric,
sorting=ynt
]{biblatex}
\AtEveryBibitem{%
    \clearfield{doi}%
  }{%
  }%

\usepackage{cleveref}
\usepackage[]{mdframed}

%% file: macros.tex
\newcommand{\leader}{\textsc{Leader}}
\newcommand{\follower}{\textsc{Follower}}

\newcommand{\eqsp}{\;}

\newcommand{\parentheseDeux}[1]{\left[ #1 \right]}
\newcommand{\parentheseDeuxRound}[1]{\left( #1 \right)}

\newcommand{\lorenzo}[1]{\todo[color=blue!40]{{\bf Lorenzo:~}#1}}

\newcommand{\paul}[1]{\todo[color=purple!40]{{\bf PM:~}#1}}

\def\rset{\mathbb{R}}

\def\nset{\mathbb{N}}

\newcommandx{\objective}[1][1=]{J_T(#1)}

\Crefname{assumNA}{{\textbf{NA}}\hspace*{-3pt}}{{\textbf{NA}}\hspace*{-3pt}}
\Crefname{assumTD}{{\textbf{TD}}\hspace*{-3pt}}{{\textbf{TD}}\hspace*{-3pt}}
\Crefname{assumFTD}{{\textbf{FTD}}\hspace*{-3pt}}{{\textbf{FTD}}\hspace*{-3pt}}
\Crefname{assumLSA}{{\textbf{LSA}}\hspace*{-3pt}}{{\textbf{LSA}}\hspace*{-3pt}}
\Crefname{assumption}{{\textbf{A}}\hspace*{-3pt}}{{\textbf{A}}\hspace*{-3pt}}

\crefname{algocf}{alg.}{algs.}
\Crefname{algocf}{Algorithm}{Algorithms}

\newcommand{\eg}{{\em e.g.,~}}

\def\rset{\ensuremath{\mathbb{R}}}
\def\nset{\ensuremath{\mathbb{N}}}

\def\PE{\mathbb{E}}

\newcommand{\beq}{\begin{equation}}
\newcommand{\eeq}{\end{equation}}
\newcommand{\beo}{\begin{array}{rl}}
\newcommand{\eeo}{\end{array}}

\def\rset{\mathbb{R}}

\def\nset{\ensuremath{\mathbb{N}}}

\def\valuefuncgen{\operatorname{V}}
\def\estvaluefunc{\hat{\valuefuncgen}}
\newcommandx{\valuefunc}[2][1=\policy_\theta,2=\discountfact]{\valuefuncgen^{#1,#2}}
\newcommandx{\statevaluefunc}[2][1=\policy_\theta,2=\discountfact]{\operatorname{Q}^{#1,#2}}
\newcommandx{\advantage}[2][1=\valuefuncgen,2=\discountfact]{\operatorname{A}^{#1,#2}}

\NewDocumentCommand{\hvaluefunc}{O{}}
{\hat{\valuefuncgen}_{#1}}
\def\policy{\pi}
\def\stateMDP{\mathsf{S}}

\def\kerMDP{\operatorname{P}}

\def\actionMDP{\mathsf{A}}
\def\rewardMDP{\bar{r}}

\def\initialMDP{\mu}

\def\actstep{\alpha}
\def\cristep{\eta}

\def\nagent{N}

\newcommandx{\statesamp}[2][1=, 2=]{S_{#1 }^{(#2)}}
\newcommandx{\actionsamp}[2][1=, 2=]{S_{#1}^{(#2)}}
\newcommandx{\nstatesamp}[2][1=, 2=]{S_{#1}^{\prime (#2)}}
\newcommandx{\mparam}[2][1=, 2=]{\param_{#1 }^{(#2)}}
\newcommandx{\mfuncAw}[2][1=, 2=]{\funcAw_{#1 }^{(#2)}}
\newcommandx{\mfuncbw}[2][1=, 2=]{\funcbw_{#1 }^{(#2)}}
\newcommandx{\mfuncA}[2][1=]{\mfuncAw[][#1](#2)}
\newcommandx{\mfuncb}[2][1=]{\mfuncbw[][#1](#2)}
\newcommandx{\mbarA}[1][1=]{\bA^{(#1)}}
\newcommandx{\mbarb}[1][1=]{\barb^{(#1)}}

\newcommandx{\fedtd}[2][1=0, 2=1]{Fed-TD($#1, #2$)\xspace}

\def\PE{\mathbb{E}}
\newcommandx{\CPE}[3][1=]{{\mathbb E}_{#1}\left[#2 \middle\vert #3 \right]} 
\newcommandx{\CPEs}[3][1=]{\PE_{#1}^{{#3}}[ #2 ]}
\newcommandx{\pscal}[3][1=]{\langle#2\,|\,#3\rangle_{#1}}
\newcommandx{\Pscal}[3][1=]{\left\langle#2\,|\,#3\right\rangle_{#1}}

\def\param{\boldsymbol{\theta}}

\DeclareMathOperator{\MSE}{MSE}
\def\bA{\bar{\mathbf{A}}}
\def\barb{\bar{\mathbf{b}}}

\def\funcAw{\mathbf{A}}

\def\funcbw{\mathbf{b}}

\def\discountfact{\gamma}

\def\msa{\mathsf{A}}

\def\mss{\mathsf{S}}

\def\rset{\mathbb{R}}

\def\nset{\mathbb{N}}

\newcommandx{\psrLigne}[3][3=]{\langle#1,#2 \rangle_{#3}}
\newcommandx{\normrLigne}[2][2=]{ \Vert#1 \Vert_{#2}}

\newcommandx{\norm}[2][1=]{\ifthenelse{\equal{#1}{}}{\left\Vert #2 \right\Vert}{\left\Vert #2 \right\Vert^{#1}}}

\newcommand\probaMarkovTilde[2][2=]
{\ifthenelse{\equal{#2}{}}{{\widetilde{\mathbb{P}}_{#1}}}{\widetilde{\mathbb{P}}_{#1}\left[ #2\right]}}

\def\eqsp{\;}

\newcommand\sequence[3][2=,3=]
{\ifthenelse{\equal{#3}{}}{\ensuremath{\{ #1_{#2}\}}}{\ensuremath{\{ #1_{#2}, \eqsp #2 \in #3 \}}}}
\newcommand\sequenceD[3][2=,3=]
{\ifthenelse{\equal{#3}{}}{\ensuremath{\{ #1_{#2}\}}}{\ensuremath{( #1)_{ #2 \in #3} }}}

\newcommand\sequenceDouble[4][3=,4=]
{\ifthenelse{\equal{#3}{}}{\ensuremath{\{ (#1_{#3},#2_{#3}) \}}}{\ensuremath{\{  (#1_{#3},#2_{#3}), \eqsp #3 \in #4 \}}}}

\newcommandx{\gperthmc}[2][1=,2=]{\ifthenelse{\equal{#1}{}}{\Xi}{\ifthenelse{\equal{#2}{}}{\Xi_{h,#1}}{\Xi_{#2,#1}}}}
\newcommandx{\Phiverlet}[2][1=,2=]{\ifthenelse{\equal{#1}{}}{\Phi}{\Phi_{#1}^{\circ (#2)}}}
\newcommandx{\gpertub}[2][1=,2=]{\ifthenelse{\equal{#1}{}}{g}{g_{#1}^{#2}}}
\newcommandx{\Phiverletq}[2][1=,2=]{\ifthenelse{\equal{#1}{}}{\widetilde{\Phi}}{\widetilde{\Phi}_{#1}^{\circ (#2)}}}
\newcommandx{\Phiverletqi}[2][1=,2=]{\ifthenelse{\equal{#1}{}}{\bar{\Psi}}{\bar{\Psi}_{#1}^{(#2)}}}
\newcommandx{\Pkerhmc}[2][1=,2=]{\ifthenelse{\equal{#1}{}}{\mathrm{P}}{\mathrm{P}_{#1, #2}}}
\newcommandx{\tPkerhmc}[2][1=,2=]{\ifthenelse{\equal{#1}{}}{\tilde{\mathrm{P}}}{\tilde{\mathrm{P}}_{#1, #2}}}
\newcommandx{\PkerhmcD}[2][1=,2=]{\ifthenelse{\equal{#1}{}}{\mathrm{K}}{\mathrm{K}_{#1, #2}}}

\DeclareMathOperator{\clip}{clip}

\def\txts{\textstyle}

%% file: introduction.tex
\section{Introduction}

V2X technology enables the bidirectional exchange of information between vehicles and other entities, such as infrastructure and pedestrians.
It relies on multiple modes of communication: vehicle-to-vehicle, vehicle-to-infrastructure, vehicle-to-pedestrian, and vehicle-to-network \cite{k1}.
The wide range of applications and the associated benefits are described in detail in \cite{k2}. 
V2X systems have strong service requirements, including minimal end-to-end latency (less than 1 ms), high data transfer rates (up to 1 Gb/s), and exceptional reliability (failure rate of less than $10^{-6}$). 
To meet these requirements, vehicles are equipped with multiple access technologies for communication in cooperative driving situations. 
Depending on their surroundings, vehicles have to decide which combination of these technologies to use, aiming to establish reliable connections, while remaining energy-efficient.
Overcoming challenges such as ensuring uninterrupted connectivity during handover and mitigating the ping-pong effect \cite{ping-pong} are thus central challenges. In this study, we propose a solution to the joint channel selection task based on Federated Deep Reinforcement Learning (FedDRL).

Many V2X tasks can be framed as sequential decision problems, making Reinforcement Learning (RL) a suitable approach. In RL, vehicles learn strategies for sequential decisions from their environment. However, RL faces challenges in real-world applications, especially in complex vehicular networks, requiring adaptation to diverse situations. FedDRL offers a solution by allowing vehicles to collaborate without sharing raw data. It extends Federated Learning (FL) to sequential decision tasks, aiming to develop global strategies applicable across all vehicles. FedDRL accelerates training by utilizing data from other vehicles and enables Deep Reinforcement Learning (DRL) models to be trained directly on edge devices like vehicles and/or roadside units. This reduces the need for extensive data transfers compared to traditional decentralized DRL \cite{barbieri2022decentralized}.
The main two contributions of this paper are the following:
\begin{itemize}
\item We develop a flexible framework for simulating federated V2X communications. Our simulator is based on Veins \cite{sommer2019veins}, and integrates with standard RL libraries. This allows for easy implementation of federated RL algorithms.
\item We apply the federated PPO methodology to joint channel selection, where multiple vehicles in a network need to communicate with each other. We demonstrate that federated PPO offers significant advantages in both learning speed and robustness, with the learned policy proving more reliable than those developed by individual agents in a non-federated setting.
\end{itemize}

%% file: related-work.tex
\section{Related Work}

RL has been applied to a variety of V2X tasks \cite{k5}, including dynamic mode selection for hybrid communication \cite{k11} and transmission power and rate selection in congestion scenarios \cite{k6}. In \cite{8633948}, the authors apply DRL-based resource allocation for V2V links. In \cite{8944302} a similar approach is used for mode selection and resource allocation in cellular V2X communication. It was also used to customize contention parameters \cite{k7} and for packet scheduling \cite{k8}.

FedDRL, an approach where multiple independent RL agents collaborate to learn how to solve a task together, has recently found many successful applications in V2X. In \cite{9511234}, FedDRL was applied to computation offloading and resource management and showed a great improvement compared to the previous state of the art. In \cite{xu2024rescale} the authors apply FedDRL to the task of resource allocation.

Several alternative approaches were investigated to get numerous agents to jointly solve a V2X task using RL:
(i) \textit{Multi-Agent Reinforcement Learning} (MARL) \cite{zhou2023multi}, where multiple agents evolve in the \emph{same} environment.
Their actions directly influence other agents, who must learn to interact with each other. MARL has been successfully applied to resource allocation in heterogeneous traffic \cite{10460728} and platooning scenarios \cite{10077432}; 
(ii) \textit{Federated Multi-Agent Reinforcement Learning}, an approach that combines MARL with FL, and has proved efficient in resource allocation problems \cite{parvini2023aoi,gui2024spectrum,9770401,xu2024rescaleinvariant}.

%% file: background.tex
\section{Federated PPO}

In this section, we describe FedPPO, a variant of PPO \cite[Algorithm 1, p.5]{schulman2017proximal} adapted to the federated setting. FedPPO allows agents to jointly learn a global policy (an actor) and a second model that estimates the quality of this policy (a critic). Both the actor and the critic are neural networks, respectively parameterized by $\theta \in \Theta \subset \mathbb{R}^{d_\theta}$ and $\phi \in \Phi \subset \mathbb{R}^{d_\phi}$.
Variants of this algorithm have already been successfully implemented in similar problems, \eg control, sensing and IoT \cite{lim2021federated, lim2020federated, ho2022federated}. 

We refer to vehicles as agents, that evolve independently from each other in their own environment. 
The dynamics of the environments are described by $N$ Markov Decision Processes (MDP). The MDP of the agent $c \in [N]$ is a tuple $(\stateMDP, \actionMDP, \kerMDP_c, \rewardMDP_c, \discountfact, \initialMDP_c)$, where
$\stateMDP$ is a continuous state space and $\actionMDP$ a finite set of actions, both common to all agents; 
for each state-action pair $(s,a) \in \stateMDP \times \actionMDP$,
$\kerMDP_c$ is a transition kernel that assigns a probability distribution $\kerMDP_c((s,a),\cdot)$ over next states, and
$\rewardMDP_c((s,a),\cdot)$ is a reward kernel, that provides the distribution of rewards; 
$\mu_c$ is the initial state distribution, and $\gamma \in [0,1]$ is the discount factor that determines the importance of future rewards.
Transition kernels 
and initial distributions
are independent, and typically differ from one agent to another.

The behavior of the agents is controlled by a shared policy $\policy_\theta: \stateMDP \times \actionMDP \rightarrow [0, 1]$, parameterized by $\theta \in \Theta \subset \mathbb{R}^{d_\theta}$, such that $\pi_{\theta}(a | s)$ specifies the probability of taking action $a$ in state $s$. The interaction between agent $c$ and its environment proceeds as follows: the agent starts in a state $s_0^{(c)}$, which is drawn from the distribution $\initialMDP_c$, then, at each time step $t$, the agent chooses an action $a_t^{(c)}$ according to the policy $\pi_{\theta}(\cdot | s_t)$. The agent then receives a reward $r_t^{(c)}$ sampled from $\rewardMDP_c((s_t^{(c)}, a_t^{(c)}), \cdot)$, and enters a new state $s_{t+1}^{(c)}$, which is determined by $\kerMDP_c(\cdot | s_t^{(c)}, a_t^{(c)})$. 
The goal of FedDRL is to find a policy that maximizes the reward obtained on average by all agents. 
This is done by maximizing the following objective function $J(\theta)$,
\begin{equation}
\label{eq:def_J}
 \txts J(\theta)= \frac{1}{N} \sum_{c=1}^N J^c(\theta) \eqsp, \text{ where } J^c(\theta) = \PE[ \sum_{t=0}^\infty \discountfact^t r_{t}^{(c)} ] \eqsp.
\end{equation}

\begin{figure}[t]
\vspace{0.25em}
    \begin{mdframed}[frametitle={Algorithm: \textsc{STEP} $(\theta_{0,0}, \phi_{0,0};
            K, B,
            \tau,
            A,
            R,
            \actstep,
            \cristep
            )$}]
    Input: 
    actor parameters $\theta_{0,0}$, 
    critic parameters $\phi_{0,0}$, 
    number of epochs $K$, 
    mini batch size $B$,
    trajectories $\tau = \{\{(s_{t}^{(m)}\!\!,a_{t}^{(m)})\}_{t=0}^T \}_{m=1}^M$,
    advantage $A = \{ \{ A_t^{(m)} \}_{t=0}^T \}_{m=1}^M$, 
    reward to-go $R = \{ \{ R_t^{(m)} \}_{t=0}^T \}_{m=1}^M$, 
    learning rate schedules $\actstep = \{ \actstep_{k}\}_{k=0}^{K-1}$,
    and $\cristep = \{ \cristep_{k}\}_{k=0}^K $.
    \begin{itemize}[leftmargin=*]
        \item 
        For $k=0, \dots K-1$ set the actor parameter $\theta_{n, k+1}$ to
        \begin{equation}
        \!\!\nonumber
            \textstyle{
            \theta_{n, k} + 
            \frac{\actstep_{k}}{B}
            \sum\nolimits_{b = 1}^{B}
            \nabla_{\theta} L(\theta_{n,k}; \theta_{n}, s_{t_{b,k}}^{( m_{b,k})}, a_{t_{b,k}}^{( m_{b,k})}, A_{t_{b,k}}^{(m_{b,k})}) \eqsp ,
            }
        \end{equation}
        with $L$ as in~\eqref{eq:ppo}, $(t_{b,k})_{1 \leq b \leq B}$ and $(m_{b,k})_{1 \leq b \leq B}$ are sets of $B$ indices drawn uniformly from $\{0,\ldots,T\}$ and $\{1,\ldots , M \}$.
        Finally, set $ \theta_{n+1} = \theta_{n,K}$ and $ \theta_{n+1,0} = \theta_{n, K}$.
    
        \item For $k=0, \dots K-1$, update the critic parameters
        \begin{equation}
        \nonumber
        \textstyle{
            \phi_{k+1}
            =
            \phi_{k} -   
            \frac{\cristep_{k}}{B}
            \sum\nolimits_{b=1}^{B}
            \nabla_{\phi} \MSE(\phi; s_{t_{b,k}'}^{(m_{b,k}')}, R_{t_{b,k}'}^{( m_{b,k}')})
            \eqsp,
            }
        \end{equation}
        with $\MSE$ as in~\eqref{eq:MSE-objective},
        $(t_{b,k}')_{1 \leq b \leq B}$ and $(m_{b,k}')_{1 \leq b \leq B}$ are sets of $B$ indices drawn uniformly from $\{0,\ldots,T\}$ and $\{1,\ldots , M \}$. 
        Finally, set $ \phi_{n+1} = \phi_{n,K}$ and $ \phi_{n+1,0} := \phi_{n,K}$.
        \end{itemize}
        Return the updated parameters $\theta_{K}, \phi_K$.
    \end{mdframed}
    \vspace{-2em}
\end{figure}

FedDRL is a framework that allows for finding the parameters that maximize this function using stochastic gradient ascent. The key feature of FedPPO is to estimate the gradient of $J$ using a surrogate objective $L(\theta) \coloneq L(\theta; \vartheta, s, a, A)$, designed to restrain policy updates using a clipping mechanism, 
\begin{equation}
        \label{eq:ppo}
         L(\theta) = \min\left( \tfrac{\pi_\theta(s, a)}{\pi_{\vartheta}(s, a)} A , \clip\left( \tfrac{\pi_\theta(s, a)}{\pi_{\vartheta}(s, a)} , 1-\epsilon, 1+\epsilon \right) A \right)
        \eqsp,
\end{equation}
where $\vartheta, \theta \in \Theta$ are estimates of the actor parameters, $(s,a) \in \mss\times \msa$ are the collected state and action, $A\in\rset$ is an estimate of the advantage related to these state and action, computed using generalized advantage estimation \cite{schulman2018highdimensional,kaledin2022variance}, and
for $x \in \rset$, the clipping operator $\clip(x, 1-\epsilon, 1+\epsilon) = \min(\max(1 - \epsilon, x), 1+\epsilon)$ (see \cite[Eq.~7, p.~3]{schulman2017proximal}) guarantees that the computed value remains within the interval $[1-\epsilon, 1+\epsilon]$. 
To estimate the parameters $\theta$ of the actor, FedPPO relies on a critic function $\estvaluefunc^{\phi}: \stateMDP \rightarrow \rset$, parameterized by $\phi$, that estimates the reward to-go $R_t$. Alike the actor, it is computed using stochastic gradient descent on the mean squared error 
\begin{align}
\textstyle
\label{eq:MSE-objective}
\MSE(\phi; s_t, R_{t})
:= ( R_{t} - \estvaluefunc^{\phi}(s_t) )^2 \eqsp ,
\end{align}
averaged over the $M$ collected trajectories $\{ \tau^{(m)} \}_{m=1}^M$.
In summary, at each communication round $n \in \nset$, FedPPO performs the following operations.
\begin{enumerate}[leftmargin=*]
    \item \textit{Local Data Collection}: for each agent $c \in \{1, \dots, N\}$, collect $M$ trajectories $\tau^{(c,n)} = \{(s_{t}^{(c,n,m)},a_{t}^{(c,n,m)}): t \in \{0,\ldots,T\}\}_{m=1}^M$ by interacting with the environment using the current policy $\pi_{\theta_n}$. 
    Compute the estimated reward-to-go $R^{(c,n)} = \{ R_{t}^{(c,n,m)}:\, t \in \{0,\ldots,T\} \}_{m=1}^M$ and generalized advantage estimators $A^{(c,n)} = \{ {A}_t^{(c,n,m)} :\, t \in \{0,\ldots,T\}\}_{m=1}^M$  with the current parameter of the critic $\phi_n$.
    \item \textit{PPO Update}: for every local iteration $h = 0, \dots, H-1$,
    \begin{itemize}[leftmargin=*]
        \item \textit{Perform local PPO Updates:} For each agent $c \in [\nagent]$, update the parameters $   \theta_{n, h+1}^{(c)}, \phi_{n, h+1}^{(c)} $ by running
        \begin{align}
            \!\!\!\!\nonumber
            \boldsymbol{ \textsc{STEP} }
                (\theta_{n, h}, \phi_{n, h};
                K, B,
                \tau_n^{(c)}\!\!,
                A^{(c,n)}\!\!,
                R^{(c,n)}\!\!,
                \actstep^{(c,n,h)}\!\!,
                \cristep^{(c,n,h)}
                )
            \eqsp,
        \end{align}   
        with $K$ local epochs, 
        batch size $B \le M T$, 
        learning rate schedule
        $\actstep^{(c,n,h)} = \{ \actstep^{(c,n,h)}_{k}\}_{k=0}^{K-1}$ for actor and 
        $\cristep^{(c,n,h)} = \{ \cristep^{(c,n,h)}_{k}\}_{k=0}^K$ for critic.

        \item \textit{Communication round}: Each agent sends its updated parameters for actor and critic networks to the central server, which aggregates them as
        \begin{equation}
        \nonumber
        \textstyle
            \theta_{n+1} = \frac{1}{N} \sum\nolimits_{c=1}^N \theta_{n,H}^{(c)}
            \eqsp, \quad \phi_{n+1} = \frac{1}{N} \sum \nolimits_{c=1}^N \phi_{n,H}^{(c)} \eqsp.
        \end{equation}
    \end{itemize}
\end{enumerate}
In the remainder of this paper, we apply this algorithm to the joint channel selection task in vehicular networks.

\begin{figure}[t]
\centering
\includegraphics[width=0.32\linewidth]{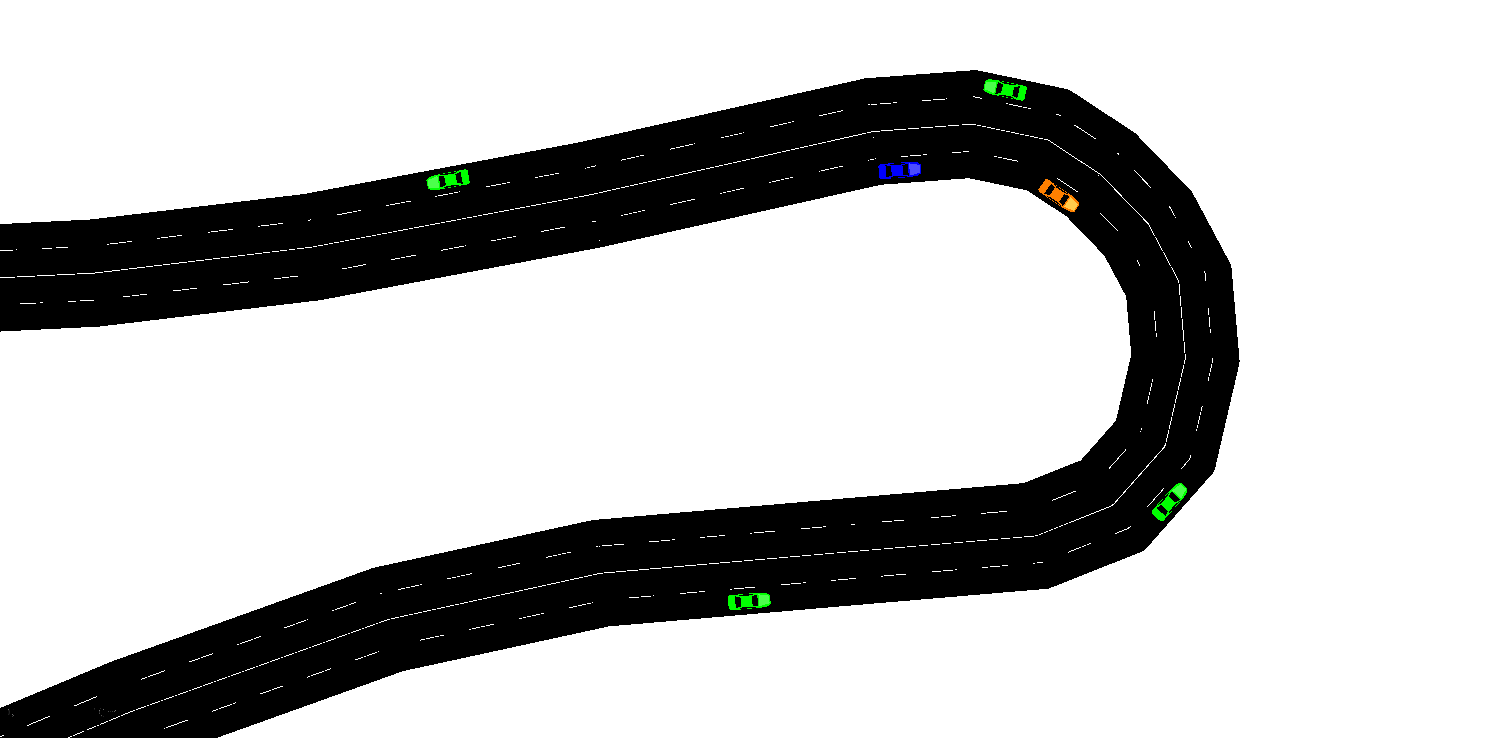}
\includegraphics[width=0.32\linewidth]{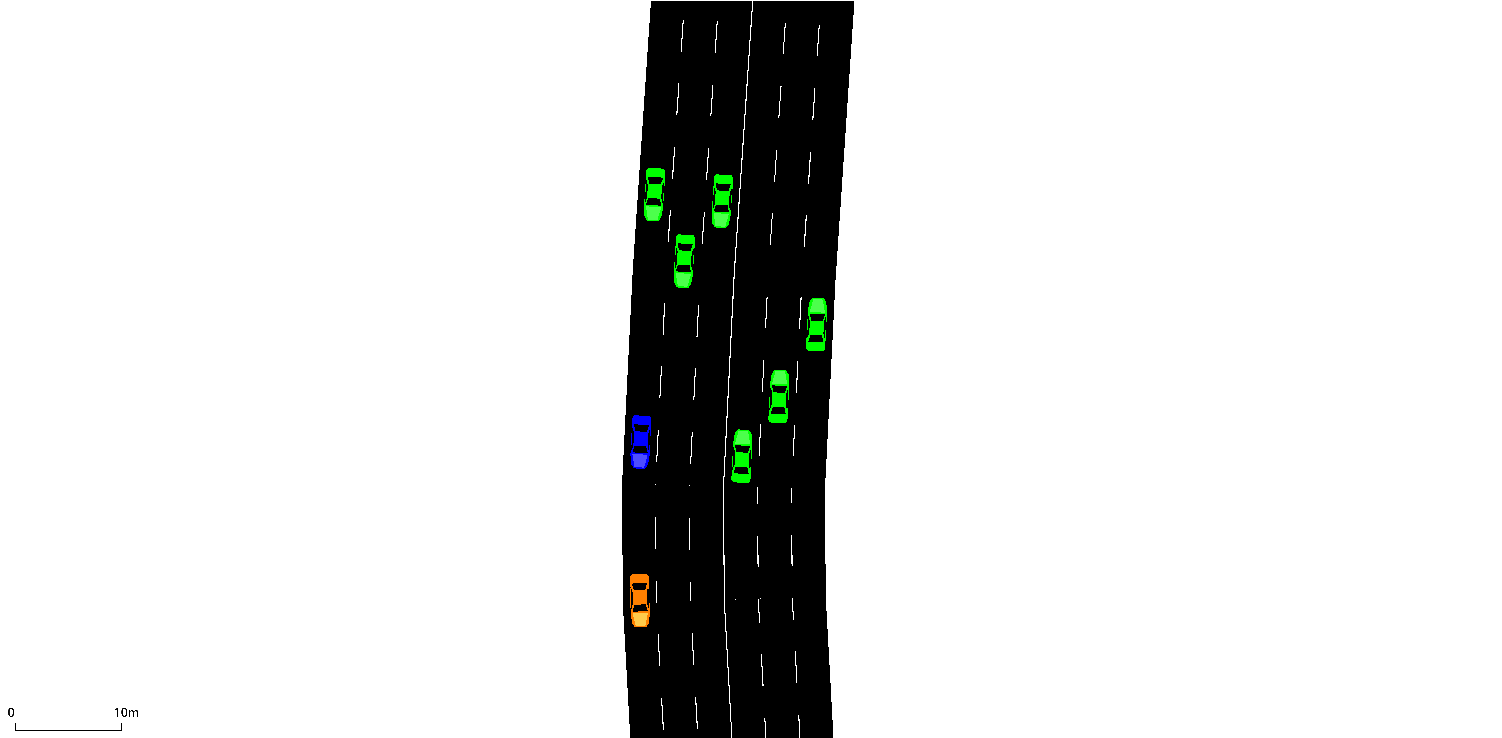}
\includegraphics[width=0.32\linewidth]{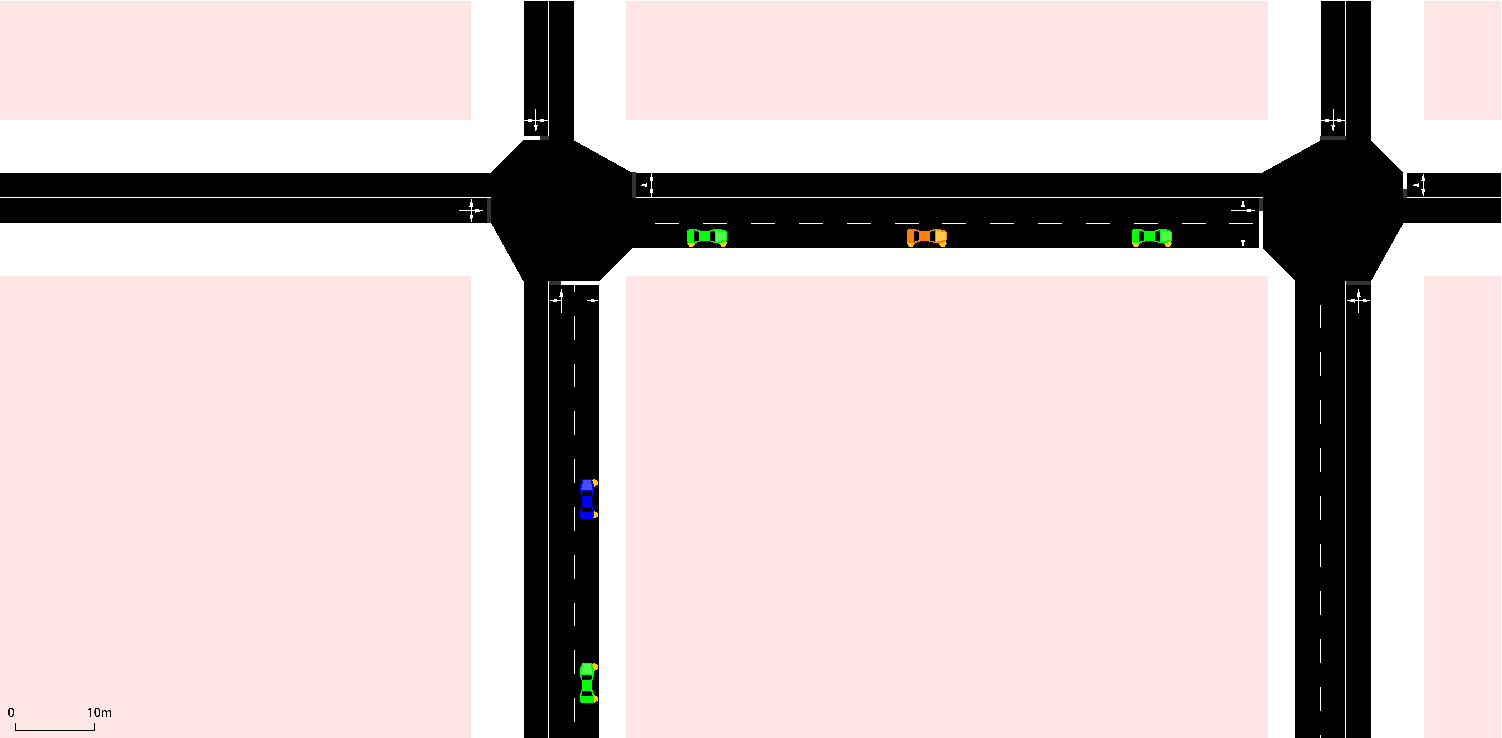}
\caption{Examples of the three traffic environments: countryside (left), highway (middle), and urban (right). The platoon involves a \follower~(blue) and a \leader~(orange), with background traffic (green). Countryside routes include multiple hairpin bends; highways are straightforward lines with good visibility; and urban routes is a grid layout with buildings that restrict the field of sight.}
\label{fig:routes}
\end{figure}

%% file: channelselection.tex
\section{FedDRL for V2X Channel Selection}
\label{sec:feddrl-in-v2x}

In this section, we apply our framework to the joint channel selection task.
We start by describing the joint channel selection problem together with the various setups of agents. Then, we evaluate the performance of FedDRL in terms of training speed and study the reliability of the learned policies.

\subsection{Channel selection use-case}
In the joint channel selection task, vehicles aim to optimize vehicle-to-vehicle communication by using a combination of access points. We consider three access points, that have to be combined to ensure reliable communication in congested roads: IEEE 802.11p Dedicated Short-Range Communications (DSRC) and Visible Light Communication with both head (VLC-H) and tail lights (VLC-T) \cite{vlc-paper}. These access points have different characteristics: DSRC is energy-intense, but allows radio communication in all directions; whereas VLC, based on visible LED light, is limited to direct line-of-sight, but consumes significantly less energy. The fundamental challenge of this channel selection task is that, in many scenarios (\eg congested roads), no single technology allows for reliable communication. Vehicles must then combine multiple access points to ensure messages are properly transmitted.

In the following, we explore scenarios where a vehicle, the \follower\ follows another one, the \leader, and aims to choose the right channels to communicate with the latter.
This problem can be formulated as an RL task, where the MDP is dictated by the road and surrounding vehicles, and the observation and actions spaces are defined as follows:
\begin{itemize}[leftmargin=*]
    \item The \emph{state} is a tuple \( s = (s_{\text{pos}}, s_{\text{net}}, s_{\text{env}}, s_{\text{action}}) \), where \( s_{\text{pos}} \) comprises five values: the relative distances between vehicles in the \( x \) and \( y \) directions, the cosine and sine of the angle of the vector pointing from the \leader\ to the \follower\ (in the coordinate system of the former), and the agent's speed. \( s_{\text{net}} \) denotes the Signal-to-Interference-plus-Noise Ratio (SINR) associated with frames received through each of the three technologies. \( s_{\text{env}} \) represents the traffic environment using one-hot encoding, with categories such as urban, countryside, and highway, as well as the placement of the DSRC antenna. \( s_{\text{action}} \) refers to the action taken at the previous time step, also encoded using one-hot encoding. The SINR is calculated by the receiving vehicle considering all incoming signals related to a specific technology. 
    
    
    \item The \emph{action} corresponds to a combination of technologies, ranging from \(0\) (no transmission) to \(7\) (all technologies): no transmission, DSRC, VLC-H, DSRC + VLC-H, VLC-T, DSRC + VLC-T, VLC-H + VLC-T, and all technologies.
    \item The \emph{reward} function is defined by the formula:
    \begin{equation*}
        r(s,a, \xi) = \xi - C(a) - \delta(s_{\text{action}}, a) \eqsp,
    \end{equation*}
where \( \xi \) takes values \( 1 \) if the message is received successfully and \( 0 \) otherwise, \( C(a) \) represents the cost associated with action \( a \) (set to \( 0.1 \) for VLC technologies and \( 0.5 \) for DSRC), and \( \delta(s_{\text{action}}, a) \) penalizes action switching: \( \delta(s_{\text{action}}, a) \) is \( 0 \) if \( a \) matches the previous action \( s_{\text{action}} \), and \( 0.01 \) otherwise.
This choice of reward is guided by the idea of maximizing successful message transmission at minimal cost while minimizing the frequency of technology switches.
\end{itemize}

\begin{table}[t]
    \centering
    \caption{Characteristics of platoon vehicles and background traffic. ``Density'' is the number of vehicles entering the road per hour.}
    \input{tables/routes}
    \label{tab:my_label}
\end{table}

\subsection{Heterogeneity}
A key asset of FedDRL is that different vehicles, that evolve in different environments, observe a more diverse part of the state space.
They evolve in heterogeneous environments, and may thus face diverse traffic environments with a variety of channel conditions. 
This allows to compensate for the typically slow progress of single-agent RL.  
In the subsequent paragraphs, we describe three different sources of heterogeneity.

\textbf{Vehicular environments.}
We use three different physical environments (\Cref{fig:routes}) that replicate specific driving conditions and interference patterns: rural, urban, and highway settings. The rural and urban environments present challenges for VLC: in rural settings, tight turns block direct light; while in urban areas, physical barriers such as buildings block the line of sight between vehicles.


\textbf{Background traffic.}
Background traffic influences communication and alters signal propagation. 
Various scenarios of background traffic can be encountered, differing in density and direction: on highways, vehicles move in both the same and opposite directions, while in rural and urban areas, they predominantly travel in opposite directions. 

\textbf{Antennas.} 
The choice of antenna significantly impacts communication channel quality. Antennas are classified based on their radiation characteristics, and each mounting option presents specific advantages and challenges affecting factors like signal obstruction, reflection, and overall coverage. We use three different antenna placements for the radio channel: a \textit{monopole} antenna mounted on the vehicle's roof, a \textit{panorama} monopole antenna positioned on the vehicle's glass roof \cite{glass-roof}, and \textit{patch} antennas mounted on the side mirrors \cite{monopole-patch}.

\subsection{Simulator}
Our federated simulation framework is based on OMNeT++ \cite{Varga2010}, Veins \cite{sommer2019veins}, and SUMO \cite{SUMO2018} to manage communication protocols in vehicular networks. OMNeT++ simulates the network and protocol development, and Veins provides models for IEEE 802.11-based communication in Vehicular Ad Hoc Networks (VANETs) and Intelligent Transportation Systems (ITS). We refer to \cite{sommer2019veins} for details on Veins. We use Veins-Gym \cite{serpentine-paper} to interface this simulator with RL algorithms, allowing to perform RL in VANET scenarios using the OpenAI Gym interface. SUMO adds realistic urban mobility scenarios to simulations, covering vehicles, bicycles, pedestrians, and more for comprehensive V2X studies.

\begin{figure}[t]
    \centering
    \includegraphics[width = 0.24\textwidth]{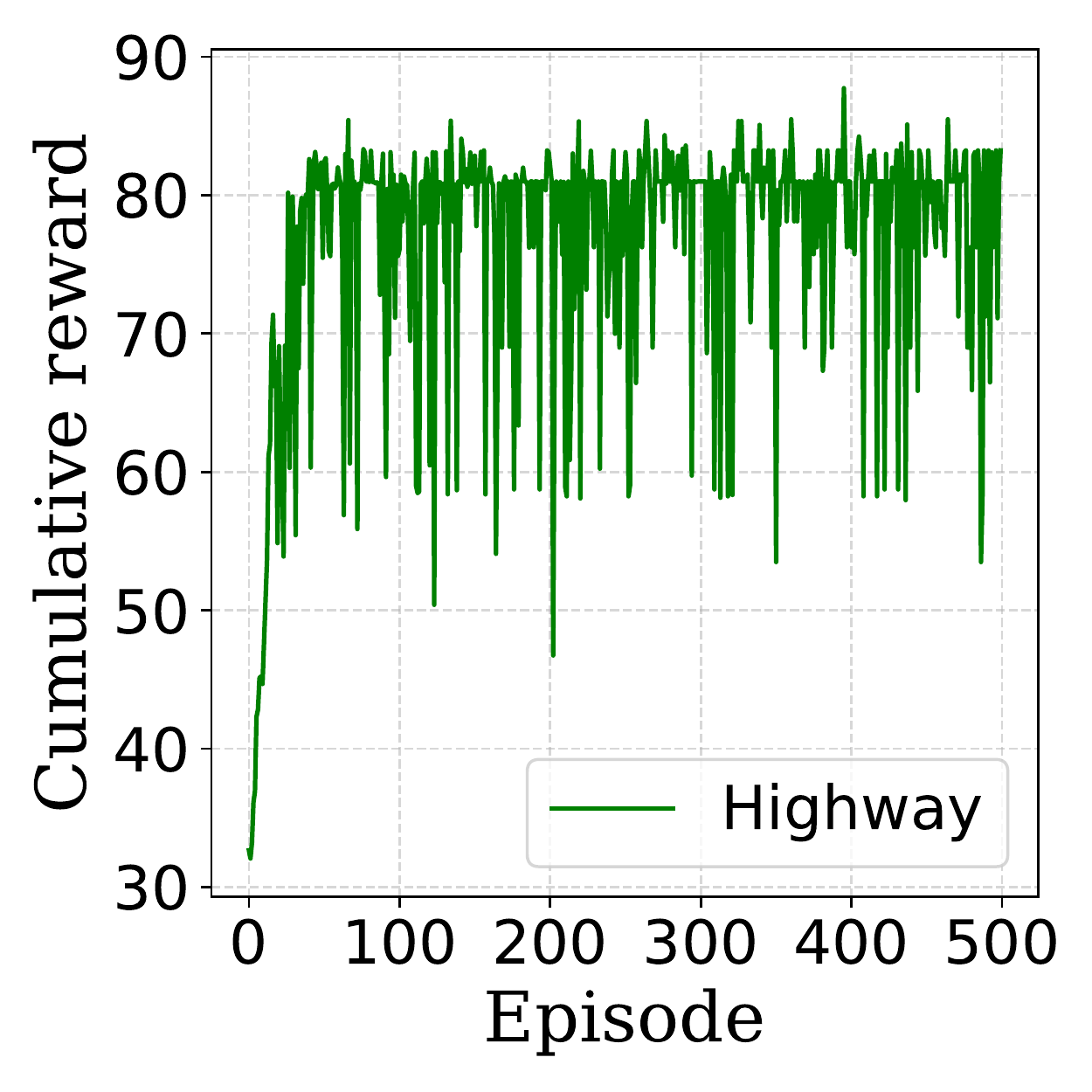}
    \includegraphics[width = 0.24\textwidth]{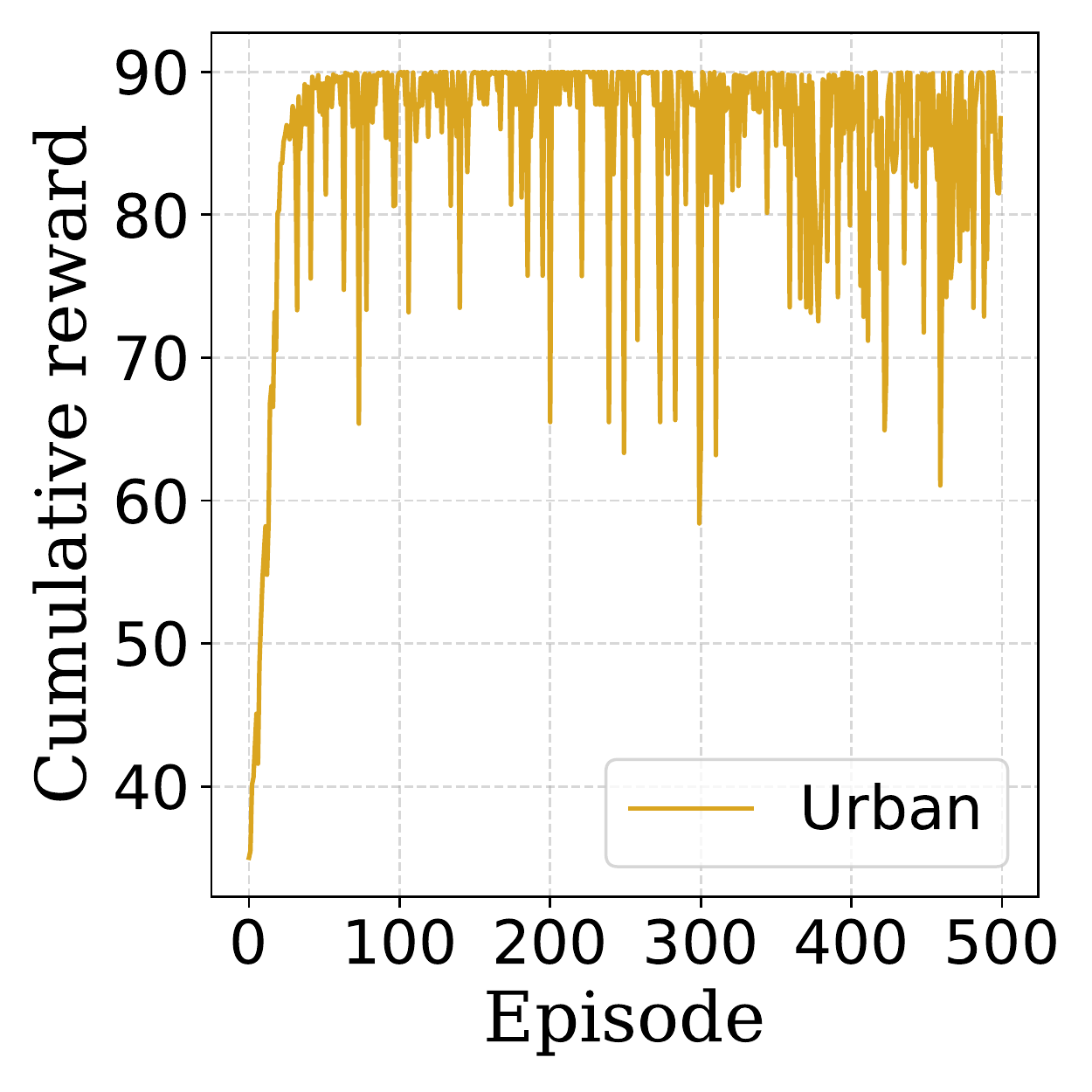}
    \includegraphics[width = 0.24\textwidth]{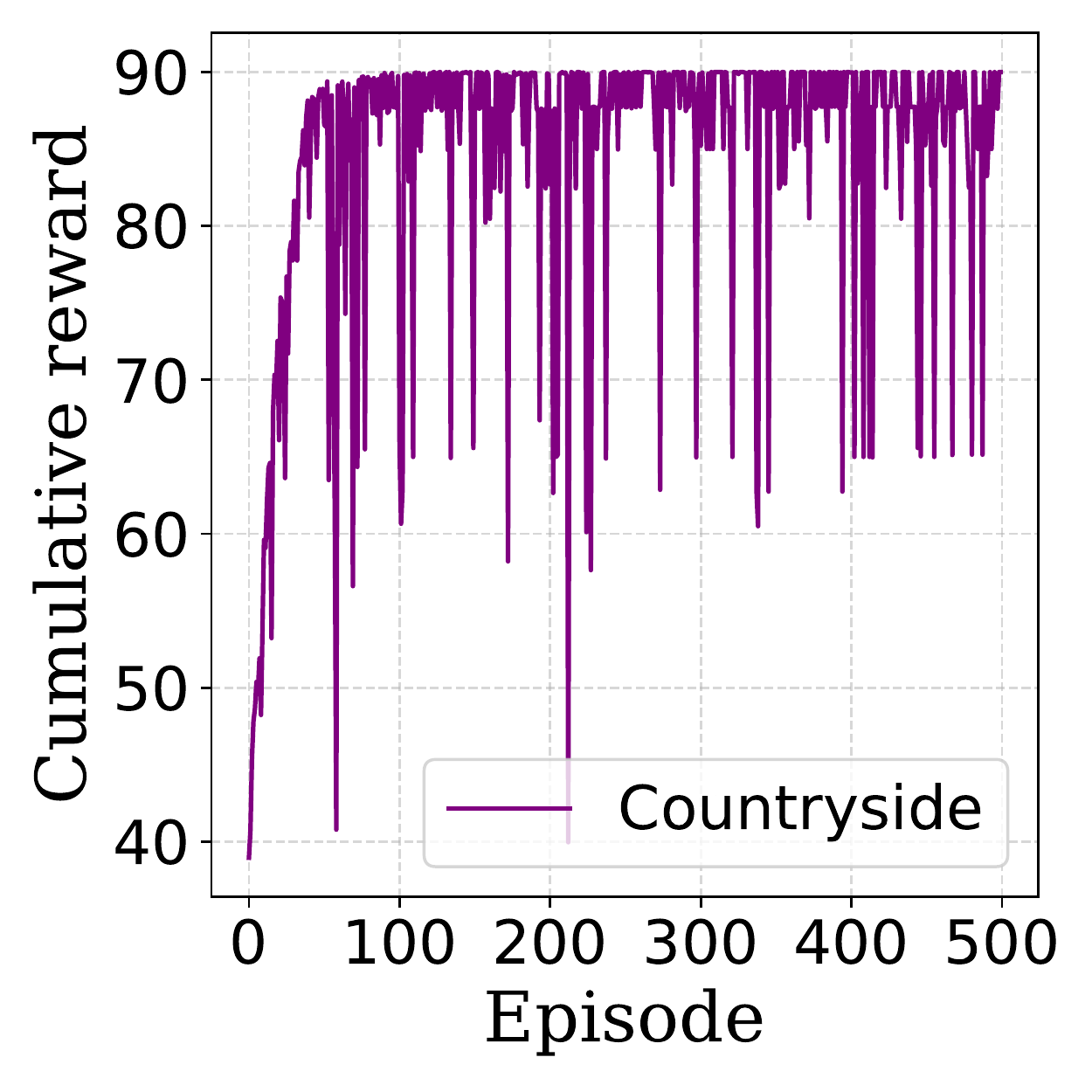}
    \includegraphics[width = 0.24\textwidth]{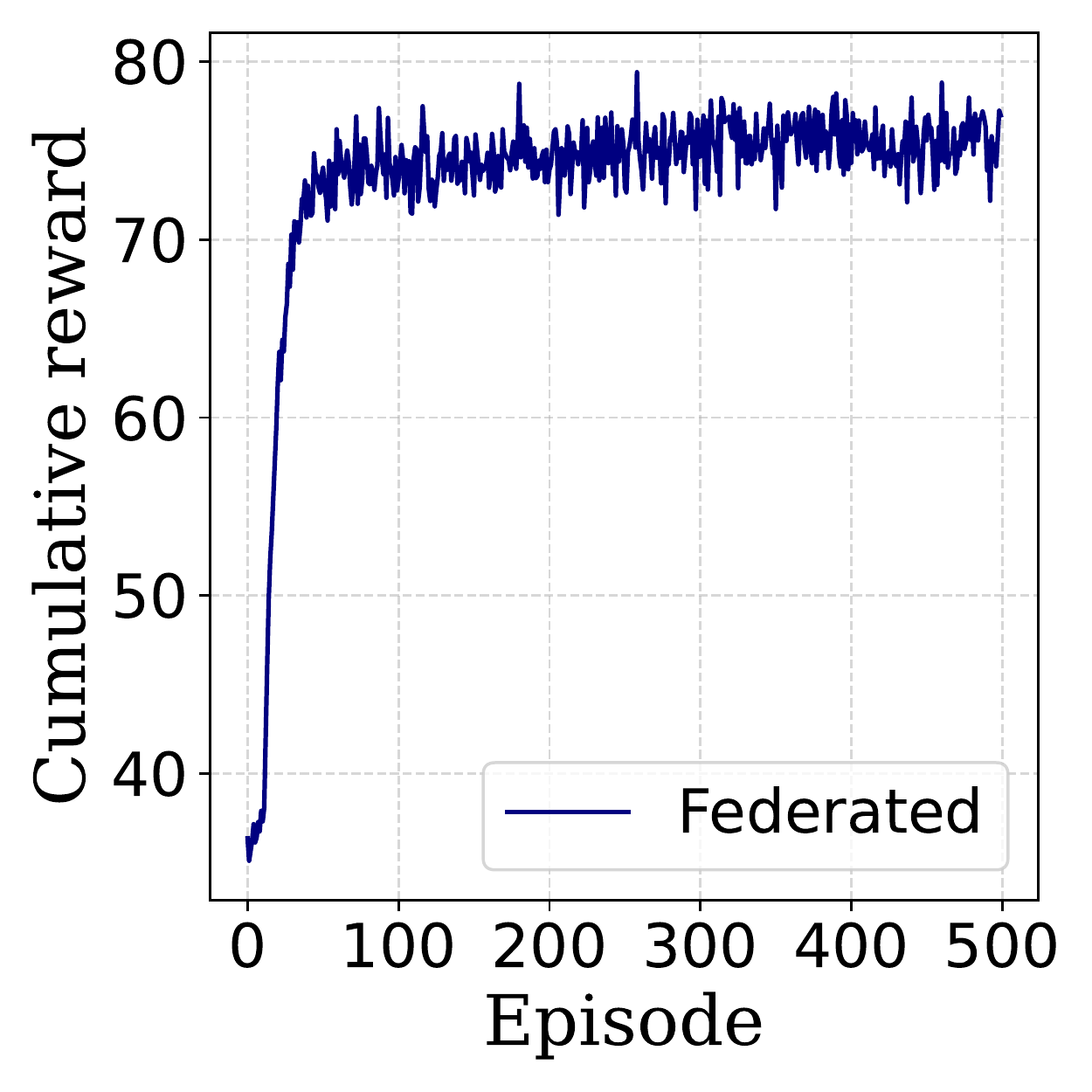}
    \caption{Cumulative rewards as a function of the number of episodes in non-federated and federated settings across three different traffic scenarios: highway, urban, and countryside.
    }
    \label{fig:learning_curve_baseline}
\end{figure}

\begin{figure}[t]
    \centering      \includegraphics[width=0.24\textwidth]{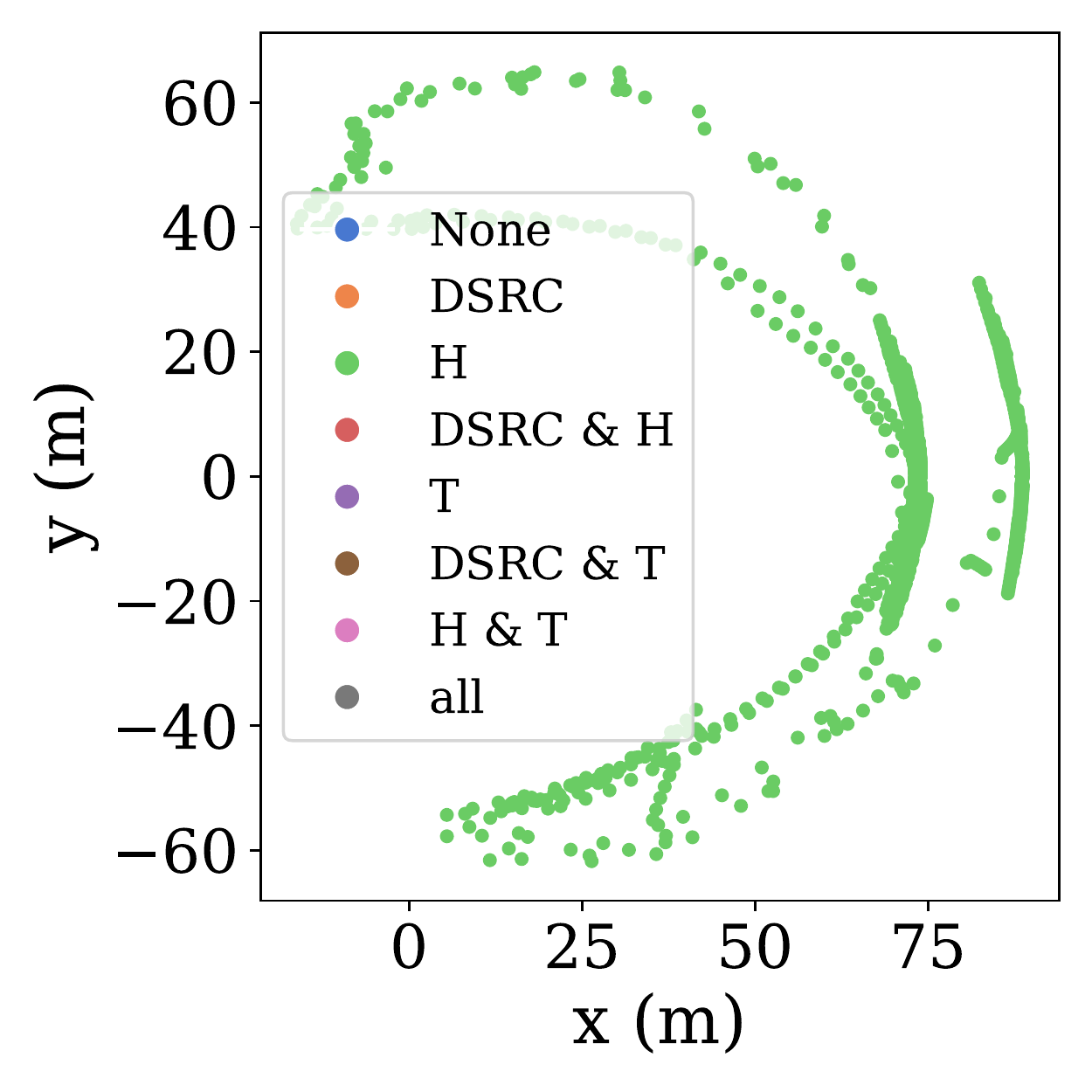}%
    \includegraphics[width=0.24\textwidth]{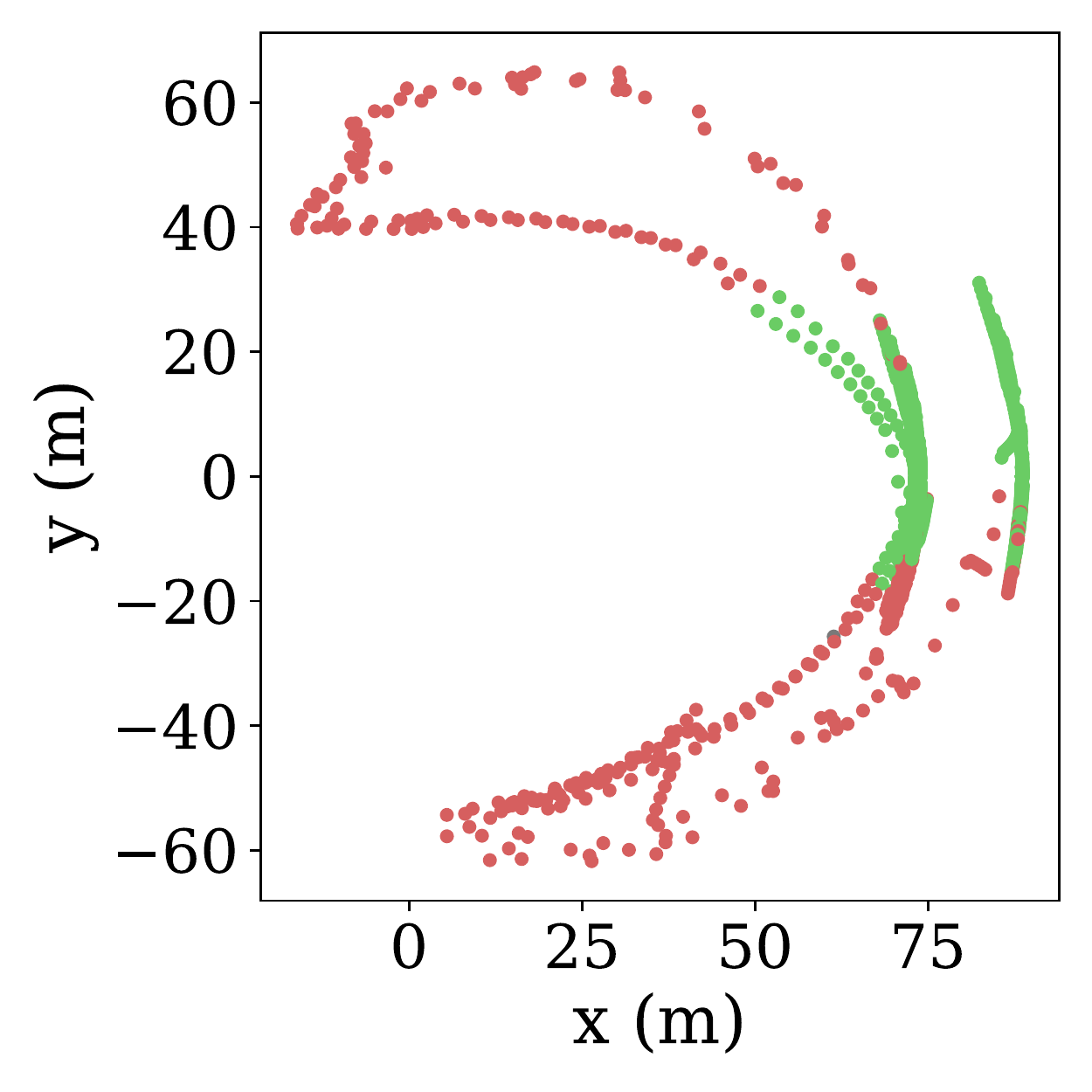}\\
    \includegraphics[width=0.24\textwidth]{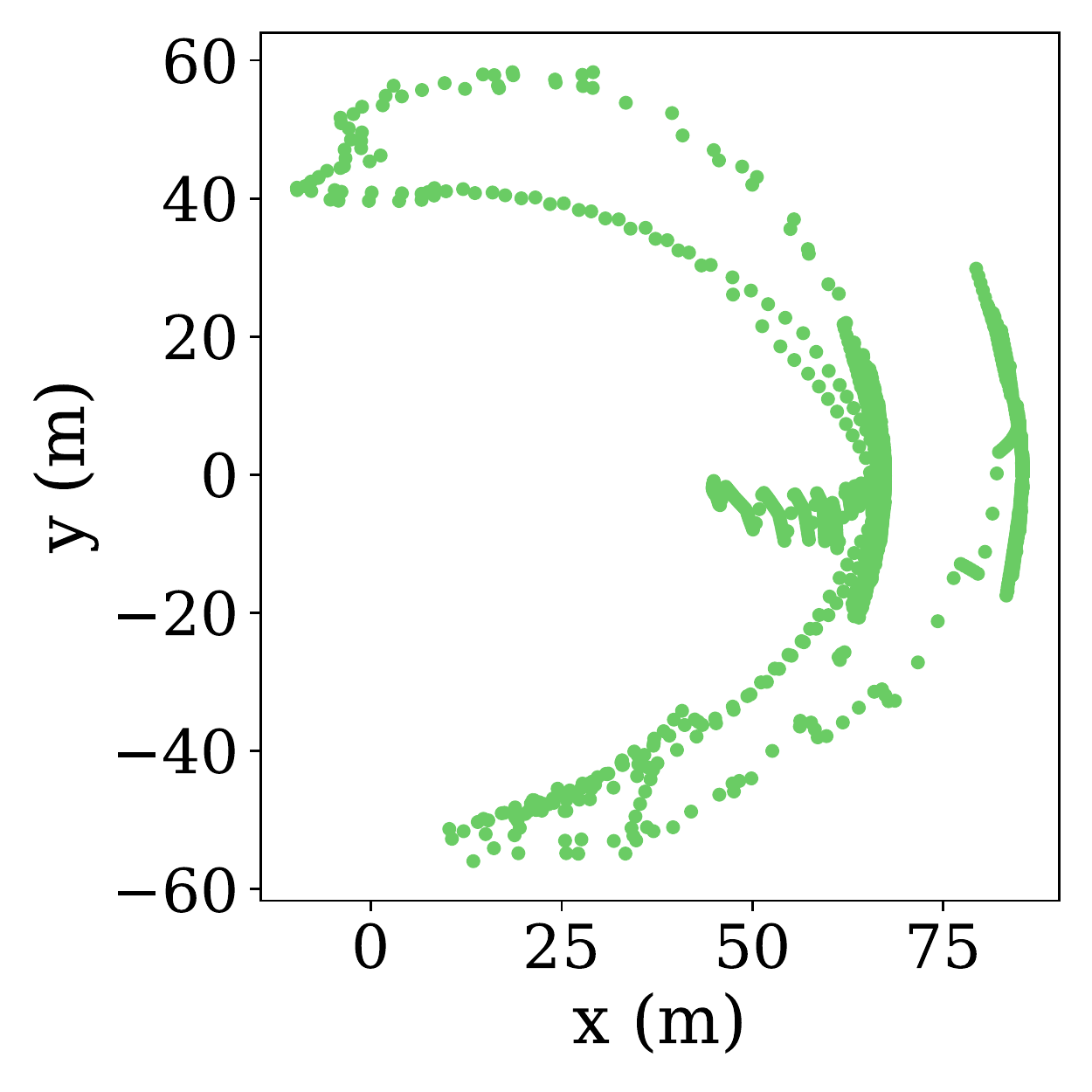}%
    \includegraphics[width=0.24\textwidth]{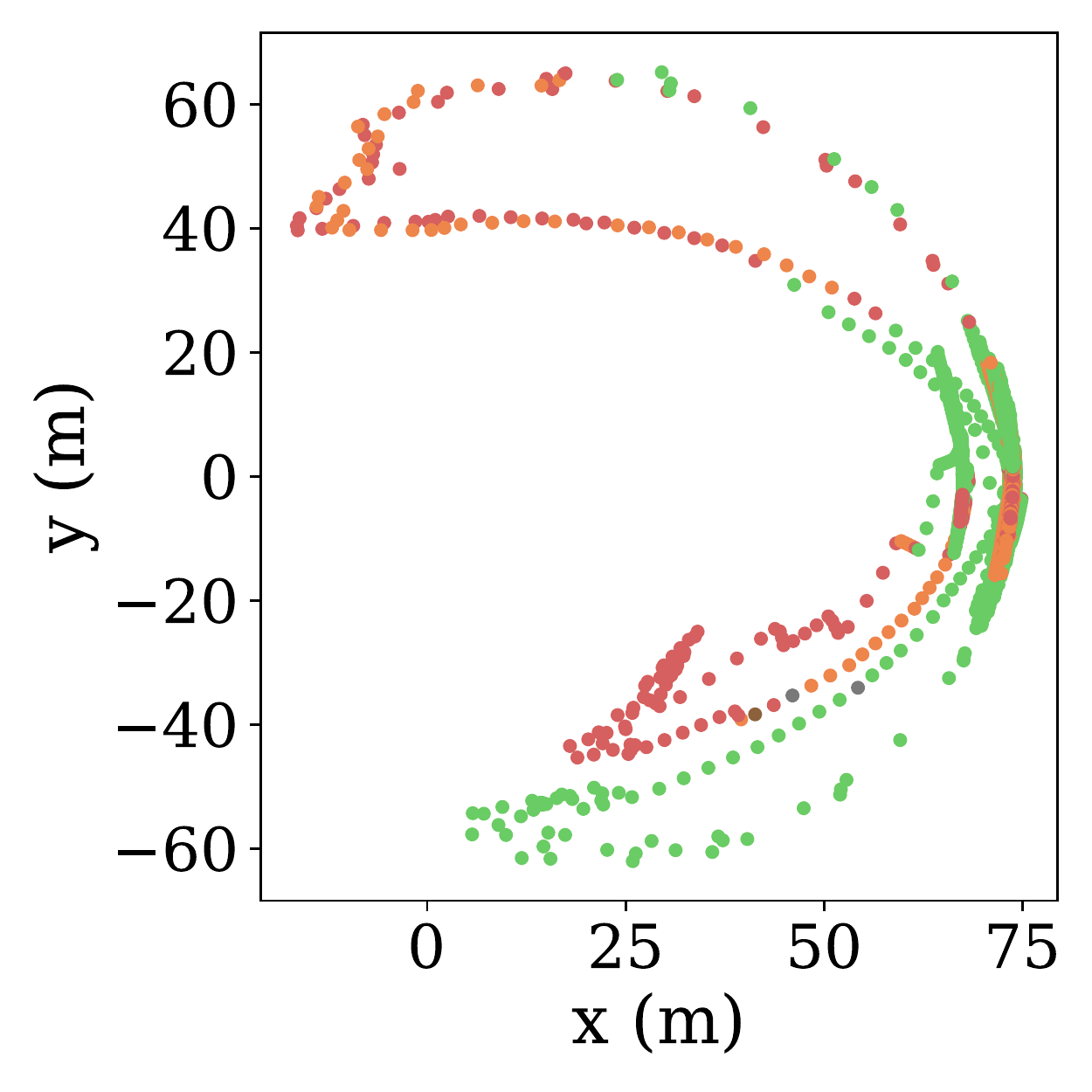}
    \caption{Decisions of the follower in the countryside environment based on the relative distance between \leader\ and the \follower. From upper left: (i) policy learned on the highway without background traffic, (ii) federated policy without background traffic, (iii) baseline policy learned on the highway with background traffic, (iv) federated policy with background traffic.}
    \label{fig:learnedstrategies}
\end{figure}

\subsection{Results}
\label{sec:results}

We now assess the applicability of FedPPO in real-world scenarios and explore the assets of FedDRL towards reliable connectivity across diverse contexts.

To serve as a reference point for comparison with our federated methodology, we start by studying the performances of three baseline scenarios. Each of these baselines consists in a single agent, only driving in one specific context. Then, agents who predominantly operate in one environment (\eg countryside), generally excel there but face difficulties in other unfamiliar contexts. Federating the training with other diverse user experiences proves advantageous, allowing agents to improve their performance across a spectrum of scenarios.
In the following, we study a practical case involving $30$ agents, evenly distributed across urban, highway, and countryside environments. Each agent predominantly operates in one setting, but occasionally encounters others, mimicking realistic driving patterns. Antenna types and background traffic levels are uniformly distributed among clients. Vehicles interact during 10-second-long simulations, during which the \follower\ selects the channels used to communicate with the \leader, while other background vehicles create interference and noise. Each vehicles conducts multiple of such episodes, allowing for the simulation of a variety of settings.

\textbf{Cumulative Rewards.}
In~\Cref{fig:learning_curve_baseline}, we show the cumulative rewards for the three centralized baselines, and for our federated methodology. 
In the three baseline scenarios, reaching a stable policy requires more iterations: the rewards remain unstable, even after several hundred episodes.
In stark contrast with this baseline, the rewards obtained by the federated vehicles are less volatile and do not exhibit significant drops.
This highlights the fact that FedDRL allows to reduce the amount of noise that is observed during the training.

\textbf{Learned Policies.} To illustrate the policies learned through FedDRL versus the non-federated baselines, we build a countryside road with four sharp corners; a setting challenging the VLC technology. Vehicles interact for about six minutes in this setup, which allows to study the learned policies in various contexts. We consider two scenarios: one with minimal background traffic and another with high traffic density.

As shown in \Cref{fig:learnedstrategies}, the FedPPO global model prioritizes VLC-H whenever possible. In narrow corners, it switches to the radio channel, often combining DSRC and VLC-H. This adjustment occurs because the algorithm needs more training to understand transmission cost differences. On the other hand, testing a policy from a single agent trained in a highway scenario reveals difficulties in corner navigation due to limited exposure of that situations in that environment.

\textbf{Reliability.} Another crucial aspect is communication reliability, measured by the Packet Delivery Ratio. As shown in \Cref{tab:reliability}, single agents perform well in their environments but struggle with different traffic contexts. In contrast, the global FedPPO model shows robust performance across all contexts. Despite this, FedPPO does not meet QoS requirements, and expanding the action space could help improve this.

\begin{table}[t]
    \centering
    \caption{Reliability of the different baselines and of the federated policy computed on the different types of environments.}
    \label{tab:reliability}
\begin{tabular}{ ccccc }
    \toprule
    \textbf{} & \textbf{Urban} & \textbf{Country} & \textbf{Highway} & \textbf{All} \\
    \midrule
    Trained on urban & $0.656$ & $0.738$ & $0.862$ & $0.752$ \\
    Trained on country & $0.654$ & $0.784$ & $0.974$ & $0.804$  \\
    Trained on highway & $0.655$ & $0.740$ & $\mathbf{0.975}$ &  $0.790$\\
    Trained on all (federated) & $\mathbf{0.724}$ & $\mathbf{0.923}$ & $0.974$ & $\mathbf{0.874}$ \\
     \bottomrule
\end{tabular}
\end{table}

%% file: tables/routes.tex
\begin{tabular}{cccc}
\toprule
        \textbf{Parameter} & 
        \textbf{Countryside}  &  \textbf{Highway} & \textbf{Urban}\\
\midrule
        Speed of vehicles     & $\sim 20 $ (m/s) &  $\sim 30 $ (m/s) & $\sim 15 $ (m/s)\\
        Background traffic density  & $0-2000$ & $0-2000$ & $0-2000$\\
\bottomrule
\end{tabular}

%% file: conclusion.tex
\section{Conclusion}

In this article, we demonstrated that the federated Proximal Policy Optimization (FedPPO) algorithm may enhance decision-making in the joint channel selection problem in V2X communications. We showed that policies learned for channel selection using FedPPO lead to better communication reliability and efficiency in V2X networks, showing that FedPPO is a promising framework for optimizing V2X communication. 
Indeed, by combining observations from multiple vehicles, FedPPO learns policies that work in a wider range of scenarios, while reducing the noise in the observed cumulative rewards when compared to non-federated approaches.

However, there are limitations affecting the performance and implementation of V2X technology with Federated Learning.
A promising direction lies in extending our simulator to include more access points (\eg 5G \cite{simu5g}, or other cellular network technologies). While our preliminary results, including DSRC and VLC, show promising results, real-life vehicles often embed many more access points, requiring to learn more complex policies.
Including additional real-world maps, modeling on existing roads, cities, and diverse traffic conditions, for instance, \cite{sumobologna}, is another promising direction. Indeed, incorporating this level of realism would provide even more accurate and meaningful insights toward improving V2X communication systems.

Overcoming these limitations is a very promising direction for further research. This would allow to further explore the potential of FedDRL in V2X applications. This perspective is even more so promising, as other FedDRL algorithms specific to V2X could be developed, integrating other baselines based on existing heuristics to further improve efficiency and reliability.

%% file: acknowledgement.tex
\section*{Acknowledgment}

The work of L.M, S.L and P.M has been supported by Technology Innovation Institute (TII), project Fed2Learn.
The work of Eric Moulines has been partly funded by the European Union (ERC-2022-SYG-OCEAN-101071601).
Views and opinions expressed are however those of the author(s) only and do not necessarily reflect those of the European Union or the European Research Council Executive Agency. Neither the European Union nor the granting authority can be held responsible for them.

%% file: bibliography.bib
@ARTICLE{8944302,
  author={Zhang, Xinran and Peng, Mugen and Yan, Shi and Sun, Yaohua},
  journal={IEEE Internet of Things Journal},
  title={Deep-Reinforcement-Learning-Based Mode Selection and Resource Allocation for Cellular V2X Communications},
  year={2020},
  volume={7},
  number={7},
  pages={6380-6391},
  doi={10.1109/JIOT.2019.2962715}}

@ARTICLE{8633948,
  author={Ye, Hao and Li, Geoffrey Ye and Juang, Biing-Hwang Fred},
  journal={IEEE Transactions on Vehicular Technology},
  title={Deep Reinforcement Learning Based Resource Allocation for V2V Communications},
  year={2019},
  volume={68},
  number={4},
  pages={3163-3173},
  doi={10.1109/TVT.2019.2897134}}

@incollection{sommer2019veins,
  author = {Sommer, Christoph and Eckhoff, David and Brummer, Alexander and Buse, Dominik S. and Hagenauer, Florian and Joerer, Stefan and Segata, Michele},
  title = {{Veins -- the open source vehicular network simulation framework}},
  editor = {Virdis, Antonio and Kirsche, Michael},
  booktitle = {Recent Advances in Network Simulation},
  doi = {10.1007/978-3-030-12842-5_6},
  isbn = {978-3-030-12841-8},
  publisher = {Springer},
  year = {2019},
}

@misc{schulman2018highdimensional,
      title={High-Dimensional Continuous Control Using Generalized Advantage Estimation},
      author={John Schulman and Philipp Moritz and Sergey Levine and Michael Jordan and Pieter Abbeel},
      year={2018},
      eprint={1506.02438},
      archivePrefix={arXiv},
      primaryClass={cs.LG}
}

@misc{kaledin2022variance,
      title={Variance Reduction for Policy-Gradient Methods via Empirical Variance Minimization},
      author={Maxim Kaledin and Alexander Golubev and Denis Belomestny},
      year={2022},
      eprint={2206.06827},
      archivePrefix={arXiv},
      primaryClass={cs.LG}
}

@misc{schulman2017proximal,
      title={Proximal Policy Optimization Algorithms},
      author={John Schulman and Filip Wolski and Prafulla Dhariwal and Alec Radford and Oleg Klimov},
      year={2017},
      eprint={1707.06347},
      archivePrefix={arXiv},
      primaryClass={cs.LG}
}

@INPROCEEDINGS{serpentine-paper,

  author={Schettler, Max and Buse, Dominik S. and Zubow, Anatolij and Dressler, Falko},

  booktitle={2020 IEEE Vehicular Networking Conference (VNC)},

  title={How to Train your ITS? Integrating Machine Learning with Vehicular Network Simulation},

  year={2020},

  volume={},

  number={},

  pages={1-4},

  doi={10.1109/VNC51378.2020.9318324}}

@INPROCEEDINGS{vlc-paper,

  author={Memedi, Agon and Tsai, Hsin-Mu and Dressler, Falko},

  booktitle={GLOBECOM 2017 - 2017 IEEE Global Communications Conference},

  title={Impact of Realistic Light Radiation Pattern on Vehicular Visible Light Communication},

  year={2017},

  volume={},

  number={},

  pages={1-6},

  doi={10.1109/GLOCOM.2017.8253979}}

@inproceedings{SUMO2018,
          title = {Microscopic Traffic Simulation using SUMO},
         author = {Pablo Alvarez Lopez and Michael Behrisch and Laura Bieker-Walz and Jakob Erdmann and Yun-Pang Fl{\"o}tter{\"o}d and Robert Hilbrich and Leonhard L{\"u}cken and Johannes Rummel and Peter Wagner and Evamarie Wie{\ss}ner},
      publisher = {IEEE},
      booktitle = {The 21st IEEE International Conference on Intelligent Transportation Systems},
           year = {2018},
        journal = {IEEE Intelligent Transportation Systems Conference (ITSC)},
            url = {https://elib.dlr.de/124092/}
 }

@ARTICLE{dsrc,

  author={Kenney, John B.},

  journal={Proceedings of the IEEE},

  title={Dedicated Short-Range Communications (DSRC) Standards in the United States},

  year={2011},

  volume={99},

  number={7},

  pages={1162-1182},

  doi={10.1109/JPROC.2011.2132790}}

@ARTICLE{k1,

  author={Zeadally, Sherali and Javed, Muhammad Awais and Hamida, Elyes Ben},

  journal={IEEE Communications Standards Magazine},

  title={Vehicular Communications for ITS: Standardization and Challenges},

  year={2020},

  volume={4},

  number={1},

  pages={11-17},

  doi={10.1109/MCOMSTD.001.1900044}}

@ARTICLE{k2,

  author={Noor-A-Rahim, Md. and Liu, Zilong and Lee, Haeyoung and Khyam, Mohammad Omar and He, Jianhua and Pesch, Dirk and Moessner, Klaus and Saad, Walid and Poor, H. Vincent},

  journal={Proceedings of the IEEE},

  title={6G for Vehicle-to-Everything (V2X) Communications: Enabling Technologies, Challenges, and Opportunities},

  year={2022},

  volume={110},

  number={6},

  pages={712-734},

  doi={10.1109/JPROC.2022.3173031}}

@ARTICLE{9770401,
  author={Li, Xiang and Lu, Lingyun and Ni, Wei and Jamalipour, Abbas and Zhang, Dalin and Du, Haifeng},
  journal={IEEE Transactions on Vehicular Technology},
  title={Federated Multi-Agent Deep Reinforcement Learning for Resource Allocation of Vehicle-to-Vehicle Communications},
  year={2022},
  volume={71},
  number={8},
  pages={8810-8824},
  doi={10.1109/TVT.2022.3173057}}

@article{gui2024spectrum,
  title={Spectrum-Energy-Efficient Mode Selection and Resource Allocation for Heterogeneous V2X Networks: A Federated Multi-Agent Deep Reinforcement Learning Approach},
  author={Gui, Jinsong and Lin, Liyan and Deng, Xiaoheng and Cai, Lin},
  journal={IEEE/ACM Transactions on Networking},
  year={2024},
  publisher={IEEE}
}

@misc{xu2024rescaleinvariant,
      title={Rescale-Invariant Federated Reinforcement Learning for Resource Allocation in V2X Networks}, 
      author={Kaidi Xu and Shenglong Zhou and Geoffrey Ye Li},
      year={2024},
      eprint={2405.01961},
      archivePrefix={arXiv},
      primaryClass={eess.SP}
}

@ARTICLE{9511234,
  author={Prathiba, Sahaya Beni and Raja, Gunasekaran and Anbalagan, Sudha and Dev, Kapal and Gurumoorthy, Sugeerthi and Sankaran, Atshaya P.},
  journal={IEEE Transactions on Network Science and Engineering},
  title={Federated Learning Empowered Computation Offloading and Resource Management in 6G-V2X},
  year={2022},
  volume={9},
  number={5},
  pages={3234-3243},
  doi={10.1109/TNSE.2021.3103124}}

@article{k5,
title = {Deep reinforcement learning techniques for vehicular networks: Recent advances and future trends towards 6G},
journal = {Vehicular Communications},
volume = {33},
pages = {100398},
year = {2022},
issn = {2214-2096},
doi = {https://doi.org/10.1016/j.vehcom.2021.100398},
url = {https://www.sciencedirect.com/science/article/pii/S221420962100067X},
author = {Abdelkader Mekrache and Abbas Bradai and Emmanuel Moulay and Samir Dawaliby}
}

@INPROCEEDINGS{k6,

  author={Roshdi, Moustafa and Bhadauria, Shubhangi and Hassan, Khaled and Fischer, Georg},

  booktitle={2021 IEEE 32nd Annual International Symposium on Personal, Indoor and Mobile Radio Communications (PIMRC)},

  title={Deep Reinforcement Learning based Congestion Control for V2X Communication},

  year={2021},

  volume={},

  number={},

  pages={1-6},

  doi={10.1109/PIMRC50174.2021.9569259}}

@misc{k7,
      title={Deep Reinforcement Learning based Joint Spectrum Allocation and Configuration Design for STAR-RIS-Assisted V2X Communications},
      author={Pyae Sone Aung and Loc X. Nguyen and Yan Kyaw Tun and Zhu Han and Choong Seon Hong},
      year={2023},
      eprint={2308.08279},
      archivePrefix={arXiv},
      primaryClass={cs.NI}
}

@ARTICLE{k8,

  author={Gu, Bo and Chen, Weixiang and Alazab, Mamoun and Tan, Xiaojun and Guizani, Mohsen},

  journal={IEEE Transactions on Vehicular Technology},

  title={Multiagent Reinforcement Learning-Based Semi-Persistent Scheduling Scheme in C-V2X Mode 4},

  year={2022},

  volume={71},

  number={11},

  pages={12044-12056},

  doi={10.1109/TVT.2022.3189019}}

@ARTICLE{k11,

  author={Yacheur, Badreddine Yacine and Ahmed, Toufik and Mosbah, Mohamed},

  journal={IEEE Transactions on Network and Service Management},

  title={Efficient DRL-Based Selection Strategy in Hybrid Vehicular Networks},

  year={2023},

  volume={20},

  number={3},

  pages={2400-2411},

  doi={10.1109/TNSM.2023.3300653}}

@Inbook{Varga2010,
author="Varga, Andras",
editor="Wehrle, Klaus
and G{\"u}ne{\c{s}}, Mesut
and Gross, James",
title="OMNeT++",
bookTitle="Modeling and Tools for Network Simulation",
year="2010",
publisher="Springer Berlin Heidelberg",
address="Berlin, Heidelberg",
pages="35--59",
isbn="978-3-642-12331-3",
doi="10.1007/978-3-642-12331-3_3",
url="https://doi.org/10.1007/978-3-642-12331-3_3"
}

@article{barbieri2022decentralized,
  title={Decentralized federated learning for extended sensing in 6G connected vehicles},
  author={Barbieri, Luca and Savazzi, Stefano and Brambilla, Mattia and Nicoli, Monica},
  journal={Vehicular Communications},
  volume={33},
  pages={100396},
  year={2022},
  publisher={Elsevier}
}

@INPROCEEDINGS{ping-pong,

  author={Kim, Won-Ik and Lee, Bong-Ju and Song, Jae-Su and Shin, Yeon-Seung and Kim, Yeong-Jin},

  booktitle={2007 IEEE 66th Vehicular Technology Conference},

  title={Ping-Pong Avoidance Algorithm for Vertical Handover in Wireless Overlay Networks},

  year={2007},

  volume={},

  number={},

  pages={1509-1512},

  doi={10.1109/VETECF.2007.321}}

@article{lim2021federated,
  title={Federated reinforcement learning acceleration method for precise control of multiple devices},
  author={Lim, Hyun-Kyo and Kim, Ju-Bong and Ullah, Ihsan and Heo, Joo-Seong and Han, Youn-Hee},
  journal={IEEE Access},
  volume={9},
  pages={76296--76306},
  year={2021},
  publisher={IEEE}
}

@article{lim2020federated,
  title={Federated reinforcement learning for training control policies on multiple IoT devices},
  author={Lim, Hyun-Kyo and Kim, Ju-Bong and Heo, Joo-Seong and Han, Youn-Hee},
  journal={Sensors},
  volume={20},
  number={5},
  pages={1359},
  year={2020},
  publisher={MDPI}
}

@article{zhou2023multi,
  title={Multi-agent reinforcement learning: Methods, applications, visionary prospects, and challenges},
  author={Zhou, Ziyuan and Liu, Guanjun and Tang, Ying},
  journal={arXiv preprint arXiv:2305.10091},
  year={2023}
}

@article{parvini2023aoi,
  title={AoI-aware resource allocation for platoon-based C-V2X networks via multi-agent multi-task reinforcement learning},
  author={Parvini, Mohammad and Javan, Mohammad Reza and Mokari, Nader and Abbasi, Bijan and Jorswieck, Eduard A},
  journal={IEEE Transactions on Vehicular Technology},
  year={2023},
  publisher={IEEE}
}

@article{ho2022federated,
  title={Federated deep reinforcement learning for task scheduling in heterogeneous autonomous robotic system},
  author={Ho, Tai Manh and Nguyen, Kim-Khoa and Cheriet, Mohamed},
  journal={IEEE Transactions on Automation Science and Engineering},
  year={2022},
  publisher={IEEE}
}

@INPROCEEDINGS{glass-roof,

  author={Kwoczek, Andreas and Raida, Zbyněk and Láčík, Jaroslav and Pokorny, Michal and Puskelý, Jan and Vágner, Petr},

  booktitle={2011 IEEE Vehicular Networking Conference (VNC)}, 

  title={Influence of car panorama glass roofs on Car2Car communication (poster)}, 

  year={2011},

  volume={},

  number={},

  pages={246-251},

  doi={10.1109/VNC.2011.6117107}}

@INPROCEEDINGS{monopole-patch,

  author={Kornek, Daniel and Schack, Moritz and Slottke, Eric and Klemp, Oliver and Rolfes, Ilona and Kürner, Thomas},

  booktitle={2010 IEEE International Conference on Communications Workshops}, 

  title={Effects of Antenna Characteristics and Placements on a Vehicle-to-Vehicle Channel Scenario}, 

  year={2010},

  volume={},

  number={},

  pages={1-5},

  doi={10.1109/ICCW.2010.5503935}}

@article{xu2024rescale,
  title={Rescale-Invariant Federated Reinforcement Learning for Resource Allocation in V2X Networks},
  author={Xu, Kaidi and Zhou, Shenglong and Li, Geoffrey Ye},
  journal={arXiv preprint arXiv:2405.01961},
  year={2024}
}

@INPROCEEDINGS{10460728,
  author={Lee, Insung and Kim, Duk Kyung},
  booktitle={2023 28th Asia Pacific Conference on Communications (APCC)}, 
  title={MARL-based Resource Allocation for Heterogeneous Traffic in V2X Communications}, 
  year={2023},
  volume={},
  number={},
  pages={61-67},
  doi={10.1109/APCC60132.2023.10460728}}

@ARTICLE{10077432,
  author={Parvini, Mohammad and Javan, Mohammad Reza and Mokari, Nader and Abbasi, Bijan and Jorswieck, Eduard A.},
  journal={IEEE Transactions on Vehicular Technology}, 
  title={AoI-Aware Resource Allocation for Platoon-Based C-V2X Networks via Multi-Agent Multi-Task Reinforcement Learning}, 
  year={2023},
  volume={72},
  number={8},
  pages={9880-9896},
  doi={10.1109/TVT.2023.3259688}}

@ARTICLE{simu5g,

  author={Nardini, Giovanni and Sabella, Dario and Stea, Giovanni and Thakkar, Purvi and Virdis, Antonio},

  journal={IEEE Access}, 

  title={Simu5G–An OMNeT++ Library for End-to-End Performance Evaluation of 5G Networks}, 

  year={2020},

  volume={8},

  number={},

  pages={181176-181191},

  doi={10.1109/ACCESS.2020.3028550}}

@inproceedings{sumobologna,
           title = {Traffic simulation for all: a real world traffic scenario from the city of Bologna},
       booktitle = {SUMO 2014},
          author = {Bieker, Laura and Krajzewicz, Daniel and Morra, Antonio Pio and Michelacci, Carlo and Cartolano, Fabio},
           month = {May},
            year = {2014},
             url = {https://elib.dlr.de/89354/}
}
